# Topological negative refraction of surface acoustic waves in a Weyl phononic crystal

Hailong He[1], Chunyin Qiu[1*], Liping Ye[1], Xiangxi Cai[1], Xiying Fan[1], Manzhu Ke[1], Fan Zhang[2], and Zhengyou Liu[1,3*]

[1]Key Laboratory of Artificial Micro- and Nano-structures of Ministry of Education and School of Physics and Technology, Wuhan University, Wuhan 430072, China

[2]Department of Physics, University of Texas at Dallas, Richardson, Texas 75080, USA

[3]Institute for Advanced Studies, Wuhan University, Wuhan 430072, China

**Reflection and refraction occur at interface between two different media. These two fundamental phenomena form the basis of fabricating various wave components. Specifically, refraction, dubbed positive refraction nowadays, appears in the opposite side of the interface normal with respect to the incidence. Negative refraction, emerging in the same side by contrast, has been observed in artificial materials[1-5] following a prediction by Veslago[6], which has stimulated many fascinating applications such as super-resolution imaging[7]. Here we report the first discovery of negative refraction of the topological surface arc states of Weyl crystals, realized for airborne sound in a novel woodpile phononic crystal. The interfaces are one-dimensional edges that separate different crystal facets. By tailoring the surface terminations of such a Weyl phononic crystal, open equifrequency contours of surface acoustic waves can be delicately designed to produce the negative refraction, to contrast the positive counterpart realized in the same sample. Strikingly different from the conventional interfacial phenomena, the unwanted reflection can be made forbidden by exploiting the open nature of the surface equifrequency contours, which is a topologically protected surface hallmark of Weyl crystals[8-12].**

Weyl semimetals[8-12], featuring doubly degenerate linear band crossing points in three-dimensional (3D) momentum space, have become a research focus in the field of topological matter[13-14]. The Weyl points are monopoles of Berry flux with topological charges characterized by quantized Chern numbers. In additional to the tantalizing bulk properties that manifest the chiral anomaly[9,15,16], the nontrivial band topology endows the Weyl semimetals with appealing surface states hosted by sample boundaries[8,10-11]. Remarkably enough, the surface band dispersion at Fermi energy forms open arcs connecting the projected Weyl points of opposite topological charges. Recently, Weyl physics has been extended to artificial structures for classical waves such as photonic crystals and phononic crystals (PCs)[17-25], in which the material ingredients with different optical or acoustic properties are periodically arranged. Soon after the experimental discovery of the photonic Weyl points in a double-gyroid structure[18], the associated surface arc states have been successfully observed in such classical systems[19-21,25]. Interestingly, the more controllable structure design and the less demanding signal detection have enabled the macroscopic classical systems to be unique platforms[26-28] for exploring the topological band physics originally proposed in electronic systems. In this Letter, based on a novel woodpile Weyl PC with delicately designed surface terminations, we present the first experimental observation of topologically protected reflectionless negative refraction of surface acoustic waves (SAWs). Our results will not only advance the current knowledge on interfacial acoustics but also enrich the Weyl physics in condensed matter systems, thereby providing a new paradigm of exotic phenomena arising from band topology.

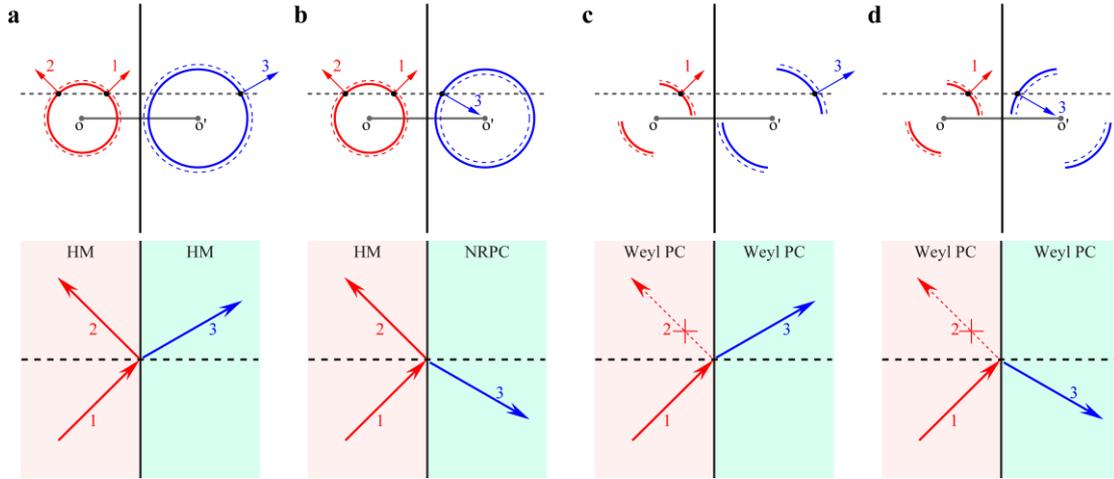

**Figure 1 | Schematics of different interfacial sound responses.** Upper and lower panels are EFC analyses and the associated beam propagations, respectively. Upper: solid curves are EFCs at the operating frequency while dashed curves are EFCs at a slightly larger frequency; circular EFCs, closed or open, are illustrated here for simplicity. Lower: the arrows 1, 2, and 3 indicate the incident, reflected, and refracted beams with their group velocities defined by the gradient vectors of EFCs in momentum space. **a**, Conventional positive refraction at an interface between two different naturally occurring homogenous media (HM). **b**, Negative refraction when the outgoing media in **a** is replaced by a PC designed with a negative refractive index (NRPC). **c**, Topologically protected positive refraction at an interface formed by two neighboring facets of a Weyl PC. **d**, The same as **c**, but for topological negative refraction. The reflection is completely forbidden in **c** and **d** due to the open EFCs, in sharp contrast to the unavoidable reflection in **a** and **b** with closed EFCs.

In general, the wave response to an interface between two different sound media can be predicted by the well-known equifrequency contour (EFC) analysis, which works for both the interfacial systems formed by naturally occurring homogenous media and/or artificially constructed periodic structures[29]. As an extension of Snell's law established in homogenous media, the EFC analysis can be made based on three fundamental criteria. The first one demands an equal frequency of the incoming and outgoing beams for the linear media considered here. The second one requires a conservation of the momentum parallel to the interface. The third one is the causality, i.e., both the reflected and refracted beams must leave the interface, yielding a constraint on their group velocities. Figure 1 illustrates various sound responses at one-dimensional (1D) or two-dimensional (2D) interfaces that separate different sound media. Comparing with the normal positive refraction (Fig. 1a) observed between two homogenous media (both featured by the EFCs expanding with frequency), an anomalous negative refraction (Fig. 1b) emerges when the outgoing medium is replaced by a PC with a negative refraction index (featured by the EFCs contracting with frequency). Note that all these EFCs are closed orbits and inevitably produce unwanted interfacial reflection. Intriguingly, both the positive and negative refractions are feasible for interfacial systems constructed using 3D Weyl PCs (Figs. 1c and 1d). Remarkably, the interfacial reflection can be made forbidden by taking the advantage of the open surface EFCs, where no reflected mode is allowed inherently. As shown below, such appealing interfacial phenomena can even be realized in a single Weyl PC with delicately designed surface terminations, whose interfaces are the 1D edges shared by two adjacent facets.

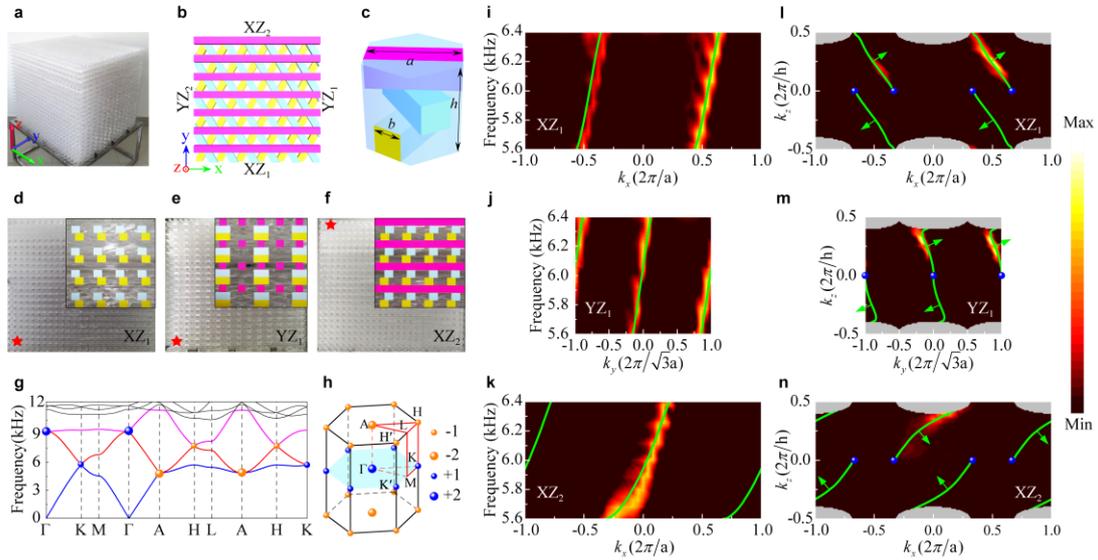

**Figure 2 | A woodpile Weyl PC and topologically protected SAWs. a**, An image of the experimental sample. **b**, Schematic top view of the trilayer-based sample. $XZ_1$, $YZ_1$, $XZ_2$, and $YZ_2$ label the four side surfaces. **c**, Geometry of the unit cell, with $a = h = 3b = 29.4$ mm. **d-f**, Front views of the three surfaces $XZ_1$, $YZ_1$, and $XZ_2$. At each surface, the red star denotes the position of a point-like sound source for

experimentally generating one-way chiral SAWs, and the colored squares and stripes are used for indicating the fine structures of the surface termination. **g**, Bulk band dispersions simulated along high-symmetry directions. **h**, The first bulk Brillouin zone of the Weyl PC. The color spheres label the Weyl points with different topological charges. **i-k**, Simulated SAW dispersions (green lines) at $k_z = 0.5\pi/h$ for the three side surfaces, precisely confirmed by our airborne sound experiments (bright color). **l-n**, The corresponding EFCs in the surface Brillouin zones, simulated and measured at the Weyl frequency 5.75 kHz. The grey regions display the projected bulk bands, the blue spheres label the projected Weyl points of 5.75 kHz, and the green arrows indicate the directions of SAW group velocities.

We consider a woodpile PC that is simply stacked by trilayer-based building blocks with a broken inversion symmetry, as shown in Figs. 2a-2c. Each trilayer unit consists of three identical square epoxy rods that are twisted anticlockwise by $2\pi/3$ along the *z* direction layer-by-layer, forming a triangular lattice in the *x-y* plane. The side length of the square rod is 9.8 mm, and both the in-plane and out-of-plane lattice constants are 29.4 mm. The experimental sample has a cuboid geometry with a size of 70.6 cm × 68.7 cm × 64.7 cm, i.e., 22 trilayer periods along the *z* direction and 24 × 27 structural periods in the *x-y* plane. Termination geometries for the four side surfaces are carefully prepared by a laser cutting technique. Note that the side surfaces $YZ_1$ and $YZ_2$ have the same appearance, whereas $XZ_1$ and $XZ_2$ are markedly different (Figs. 2b and 2d-2f). Such a design lends the SAW EFCs desired properties for realizing the topological negative refraction (Fig. 1d) at one edge, together with the comparative positive refraction (Fig. 1c) at another edge of the same sample. As illustrated in Figs. 2d-2f, three connected side surfaces, $XZ_1$, $YZ_1$, and $XZ_2$, are used for demonstrating such intriguing interfacial phenomena. In order to best characterize the one-way chiral SAWs and their interfacial responses, the three side surfaces are casted onto the same plane, and a unified *x* axis pointing to the right is used for the surfaces $XZ_1$ and $XZ_2$. Acoustically hard epoxy plates, as trivial insulators for sound, are closely attached on the sample to host the topologically protected SAWs.

Figure 2g shows the bulk band dispersions for the Weyl PC along directions of high symmetry. Interestingly, this structurally simple PC possesses not only single Weyl points but also double Weyl points, which link together the lowest three bands (color lines). The single Weyl points, around which the dispersions are linear along all directions, emerge at K and H points that are not invariant under time reversal, whereas the double Weyl points, around which the dispersions become quadratic in the $k_x$-$k_y$ plane, locate at Γ and A points as dictated by the time-reversal symmetry. The stability of these Weyl points at the high-symmetry momenta is a direct consequence of the threefold screw symmetry of the Weyl PC. The topological charges of these Weyl points, as sources or sinks of Berry flux, can be identified by analyzing the rotational eigenvalues at their high-symmetry momenta[20,30]. These topological charges and their distribution (Figs. 2g and 2h) lead to topologically nontrivial SAWs on the truncated surfaces parallel to the *z* direction. This is demonstrated in Figs. 2i-2k by the numerical SAW dispersions (green lines) simulated at $k_z = 0.5\pi/h$ for the XZ$_1$, YZ$_1$, and XZ$_2$ surfaces, which traverse a wide gap between the lowest two bulk bands. Because all these dispersions have positive slopes, the $k_z$-fixed SAWs propagate anticlockwise around the sample (viewed from the top). This one-way chiral behavior of the SAWs can be understood by the fact that the spiral interlayer coupling in the woodpile PC introduces a synthetic gauge flux threading through the sample along the *z* direction. Note that the surface conditions greatly influence the detailed properties of the SAW dispersions hosted by the XZ$_1$ and XZ$_2$ surfaces (Figs. 2i and 2k). This is further demonstrated in Figs. 2l and 2n by the open EFCs (green lines) simulated at 5.75 kHz, the frequency of the Weyl points with topological charge +1. Specifically, for $k_z > 0$ the SAWs at the XZ$_2$ surface always carry a downward component of the group velocity, in contrast to the upward one at the XZ$_1$ surface. (For the SAWs with $k_z < 0$, the group velocities reverse automatically due to time-reversal symmetry.) Such contrasting features contribute significantly to the distinct SAW responses to the two 1D edges of YZ$_1$ surface shared with XZ$_1$ and XZ$_2$ surfaces. In addition, these surface EFCs exhibit very gentle curvatures in a wide range of momenta, and the majority of SAWs propagate toward nearly the same direction. This facilitates the experimental observation of the fascinating positive and negative refraction of the topologically protected SAWs.

The presence of the topologically nontrivial SAWs is validated by our airborne sound experiments. A broadband point-like sound generator is positioned under the cover plate to excite the SAWs, and a subwavelength-sized sound probe is inserted inside the sample to scan the surface pressure signal point by point. Particularly, in each sample surface the sound source is placed on a specific surface corner (Figs. 2d-2f) to stimulate the one-way chiral SAWs. In this way, all SAWs that propagate rightwards with $k_z > 0$ are selectively excited. This not only reduces the finite size effect but also silence the information of the time-reversal counterparts (i.e., the SAWs with $k_z < 0$). By Fourier transformation of the near-field pressure distributions, we map out the SAW dispersions in the surface Brillouin zones. Figures 2i-2k present the experimental SAW dispersions (color) in the $k_x$-$k_y$ plane for $k_z = 0.5\pi/h$, and Figs. 2l-2n display the measured surface EFCs at the Weyl frequency 5.75 kHz. Excellent agreements can be found between the experiments (bright color) and simulations (green lines) for all the three surfaces, in spite of the band broadening due to the finite size effect.

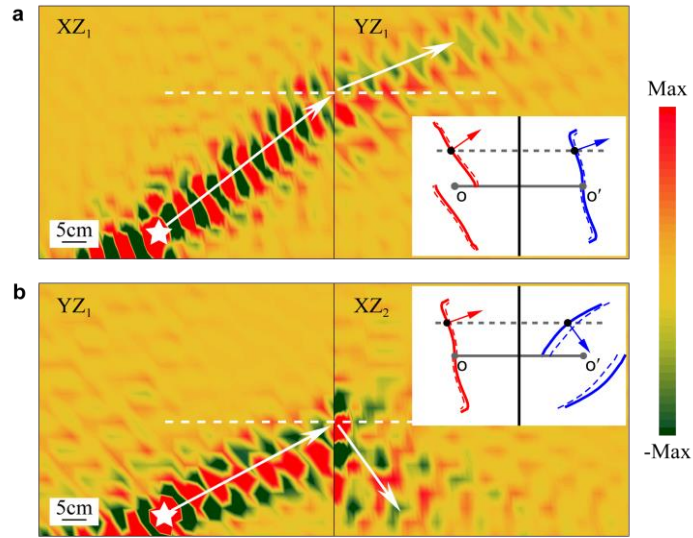

**Figure 3 | Experimental observation of the topological negative refraction. a,b**, Positive and negative SAW responses to the two 1D edges of $YZ_1$ surface shared with $XZ_1$ and $XZ_2$ surfaces. The near-field pressure distributions are measured at the frequency 5.75 kHz. The white stars label the positions of the point-like sound sources, and the white arrows indicate the directions of the propagating beams. Insets: Schematics of the echo-free interfacial phenomena, where the solid and dashed curves depict the EFCs at 5.75 kHz and 5.80 kHz, respectively. Both cases exhibit negligible interfacial reflections.

Now we turn to the experiments of interfacial responses of the topologically protected SAWs. Two comparative experiments are carried out, one for the interface between the $XZ_1$ and $YZ_1$ surfaces, and the other for the interface between the $YZ_1$ and $XZ_2$ surfaces. Figure 3 shows the measured near-field pressure distributions at 5.75 kHz. In each case, the point-like sound source is positioned at the bottom of the initial surface ($XZ_1$ or $YZ_1$) to excite all SAWs with $k_z > 0$, which propagate uniformly toward the upper right. As anticipated from the EFC analysis (inset in Fig. 3a) introduced above, Fig. 3a displays a typical positive refraction effect when the sound signal traverses across the interface to reach the second surface $YZ_1$, where the refracted beam travels upward and deflects in the opposite side of the interface normal (white dashed line) with respect to the incoming beam. The finite widths of the incoming and outgoing beams in real space stem mostly from the finite lengths of the open EFCs with $k_z > 0$. By striking contrast, Fig. 3b shows a negative refraction since the outgoing beam propagates downward and deflects in the same side of the interface normal. This is consistent with the fact that the *z*-component of the group velocity is reversed for the SAWs from the YZ$_1$ to XZ$_2$ surfaces (inset in Fig. 3b). In particular, for each interfacial system the initial beam crosses the 1D interface smoothly to the neighboring surface and exhibits no observable reflection signal, unlike the conventional wave response to an interface. Physically, these echo-free interfacial phenomena originate from the topological nature of the SAWs that are robust against any $k_z$-preserving scattering, even though the surfaces bend sharply near the edges (90$^o$ for both cases here).

In conclusion, we have fabricated a novel Weyl PC and shown a direct experimental evidence for the topological negative refraction of SAWs. It is worth pointing out that the flexibility in tailoring the open EFCs of SAWs through engineering the sample terminations enables a wide variety of exceptional interfacial responses to sound, such as focusing and imaging without any reflection. Such phenomena should also be observable in other Weyl systems and may offer unprecedented possibilities for controlling waves.

**Methods**

**Numerical simulations.**

All full-wave simulations are performed by a commercial solver package (COMSOL Multiphysics). The epoxy material used in our experiments is safely modeled as an acoustically rigid material, given the great impedance mismatch with respect to the air background. The airborne sound speed 346 m/s is used for our experiment at room temperature 23$^o$C. The bulk band dispersions (Fig. 2g) are calculated by a single unit cell (Fig. 2c) with Bloch boundary conditions in all 3D directions. To calculate the $k_z$-fixed surface band dispersions (Figs. 2i-2k), we consider infinitely large slab structures with terminations and thicknesses specified in the text. In each case, the slab is thick enough to avoid the coupling between the SAWs hosted by the two different surfaces. Rigid boundary conditions are applied to both slab surfaces, whereas Bloch boundary conditions are used for the remaining two directions, with fixed $k_z$ and varying $k_x$ or $k_y$. In addition to the projected bulk bands, each numerical configuration gives the SAWs for both surfaces simultaneously, which can be further distinguished by inspecting the surface localization of the field distributions. Similarly, the EFCs (Figs. 2l-2n) are extracted by scanning the surface Brillouin zone at the frequency 5.75 kHz. In addition to the numerical data provided in Fig. 2, more simulations about the bulk and surface dispersions can be found in our supplemental material.

**Experiment measurements.**

To experimentally excite the SAWs, a broadband sound signal is launched from a narrow tube (of radius ~4.0 mm) that penetrates the epoxy plate covering the Weyl PC sample. The sound source behaves point-like for the wavelength ~60 mm focused here. The surface field is scanned point-by-point through a microphone (of radius ~3.5 mm, B&K Type 4187) that is inserted inside the sample. The field confinement of the topologically protected SAWs is checked by detecting the pressure distributions away from the surface (see supplemental material). The scanning steps are 5.0 mm, 25.4 mm, and 29.4 mm along the *x*, *y*, and *z* directions, respectively. Both the amplitude and phase information of the local pressure field can be recorded and frequency-resolved by a multi-analyzer system (B&K Type 3560B). To map out the EFCs (Figs. 2l-2n) in the surface Brillouin zone, 2D Fourier transformation is performed for the measured spatial pressure distributions at a given frequency. This further gives the frequency dependent dispersion curves for a fixed $k_z$ (Figs. 2i-2k).

**References**

1. Shelby, R. A., Smith, D. R. & Schultz, S. Experimental verification of a negative index of refraction. *Science* **292**, 77-79 (2001).

2. Cubukcu, E., Aydin, K., Ozbay, E., Foteinopoulou, S. & Soukoulis. C. M. Electromagnetic waves: Negative refraction by photonic crystals. *Nature* **423**, 604-605 (2003).

3. Yao, J. *et al.* Optical negative refraction in bulk metamaterials of nanowires. *Science* **321**, 930 (2008).

4. Yang, S. *et al.* Focusing of sound in a 3D phononic crystal. *Phys. Rev. Lett.* **93**, 024301 (2004).

5. Zhang, S., Yin, L. & Fang, N. Focusing ultrasound with an acoustic metamaterial network, *Phys. Rev. Lett.* **102**, 194301 (2009).

6. Veselago, V. G. The electrodynamics of substances with simultaneously negative values of $\varepsilon$ and $\mu$. *Sov. Phys. Usp.* **10**, 509 (1968).

7. Pendry, J. B. Negative Refraction Makes a Perfect Lens. *Phys. Rev. Lett.* **85**, 3966 (2000).

8. Wan, X., Turner, A. M., Vishwanath, A. & Savrasov, S. Y. Topological semimetal and Fermi-arc surface states in the electronic structure of pyrochlore iridates. *Phys. Rev. B* **83**, 205101 (2011).

9. Burkov, A. A. & Balents, L. Weyl semimetal in a topological insulator multilayer. *Phys. Rev. Lett.* **107**, 127205 (2011).

10. Xu, S.-Y. *et al.* Discovery of a Weyl fermion semimetal and topological Fermi arcs. *Science* **349**, 613-617 (2015).

11. Lv, B. Q. *et al.* Experimental discovery of Weyl semimetal TaAs. *Phys. Rev. X* **5**, 031013 (2015).

12. Soluyanov, A. A. *et al.* Type-II Weyl semimetals. *Nature* **527**, 495-498 (2015).

13. Hasan, M. Z. & Kane, C. L. Colloquium: topological insulators. *Rev. Mod. Phys.* **82**, 3045-3067 (2010).

14. Qi, X.-L. & Zhang, S.-C. Topological insulators and superconductors. *Rev. Mod. Phys.* **83**, 1057-1110 (2011).

15. Xu, G., Weng, H., Wang, Z., Dai, X. & Fang, Z. Chern semimetal and the quantized anomalous hall effect in HgCr2Se4. *Phys. Rev. Lett.* **107**, 186806 (2011).

16. Huang, X. *et al.* Observation of the chiral-anomaly-induced negative magnetoresistance in 3D Weyl semimetal TaAs. *Phys. Rev. X* **5**, 031023 (2015).

17. Lu, L., Fu, L., Joannopoulos, J. D. & Soljacic, M. Weyl points and line nodes in gyroid photonic crystals. *Nature Photon.* **7**, 294-299 (2013).

18. Lu, L. *et al.* Experimental observation of Weyl points. *Science* **349**, 622-624 (2015).

19. Chen, W. J., Xiao, M. & Chan, C. T. Photonic crystals possessing multiple Weyl points and the experimental observation of robust surface states. *Nature Commun.* **7**, 13038 (2016).

20. Noh, J. *et al.* Experimental observation of optical Weyl points and Fermi arc-like surface states. *Nat. Phys.* **13**, 611-617 (2017).

21. Yang, B. *et al*. Direct observation of topological surface-state arcs in photonic metamaterials, *Nat. Commun.* DOI: 10.1038/s41467-017-00134-1.

22. Chang, M. L., Xiao, M., Chen, W. J. & Chan, C. T. Multiple Weyl points and the sign change of their topological charges in woodpile photonic crystals. *Phys. Rev. B* **95**, 125136 (2017).

23. Xiao, M., Chen, W. J., He, W. Y. & Chan, C. T. Synthetic gauge flux and Weyl points in acoustic systems. *Nat. Phys.* **11,** 920-924 (2015).

24. Yang, Z. & Zhang, B. Acoustic type-II Weyl nodes from stacking dimerized chains. *Phys. Rev. Lett.* **117**, 224301 (2016).

25. Li, F., Huang, X., Lu, J., Ma, J., & Liu, Z. Weyl points and Fermi arcs in a chiral phononic crystal. *Nat. Phys.* DOI: 10.1038/NPHYS4275.

26. Wang, Z., Chong, Y., Joannopoulos, J. D.& Soljacic, M. Observation of unidirectional backscattering-immune topological electromagnetic states. *Nature* **461**, 772-775 (2009).

27. Rechtsman, M. C. *et al*. Photonic floquet topological insulators. *Nature* **496**, 196-200 (2013).

28. Susstrunk, R. & Huber, S. D. Observation of phononic helical edge states in a mechanical topological insulator. *Science* **349,** 47-50 (2015).

29. Notomi, M. Theory of light propagation in strongly modulated photonic crystals: Refractionlike behavior in the vicinity of the photonic band gap. *Phys. Rev. B* **62**, 10696 (2000).

30. Fang, C., Gilbert, M. J., Dai, X. & Bernevig, B. A. Multi-Weyl Topological Semimetals Stabilized by Point Group Symmetry. *Phys. Rev. Lett.* **108**, 266802 (2012).

**Acknowledgements**

This work is supported by the National Basic Research Program of China (Grant No. 2015CB755500); National Natural Science Foundation of China (Grant Nos. 11774275, 11674250, 11534013, 11547310); Natural Science Foundation of Hubei Province (Grant No. 2017CFA042). FZ was supported by the UT-Dallas research enhancement funds.

**Author contributions**

C.Q. and Z.L. conceived the original idea and supervised the project. H.H. performed the simulations. H.H., L.Y., X.F. and M.K. carried out the experiments. C.Q., H.H., X.C., F.Z. and Z.L. analyzed the data and wrote the manuscript. All authors contributed to scientific discussions of the manuscript.

**Additional information**

Supplementary information is available in the online version of the paper. Reprints and permission information are available online at www.nature.com/reprints. Correspondence and requests for materials should be addressed to C.Q. and Z. L.

**Competing Interests**

The authors declare that they have no competing financial interests.